\documentclass[twocolumn, amsmath, amssymb, aps, prl, superscriptaddress]{revtex4-2}
\usepackage{xr}
\usepackage{amsmath}
\usepackage{verbatim}

\makeatletter
\newcommand*{\addFileDependency}[1]{
  \typeout{(#1)}
  \@addtofilelist{#1}
  \IfFileExists{#1}{}{\typeout{No file #1.}}
}
\makeatother
\usepackage{lineno}

\newcommand*{\myexternaldocument}[1]{%
    \externaldocument{#1}%
    \addFileDependency{#1.tex}%
    \addFileDependency{#1.aux}%
}
\myexternaldocument{Supplemental}

\usepackage{graphicx}   
\usepackage{dcolumn}  
\usepackage{bm}          
\usepackage{xcolor}    

\usepackage{hyperref}
\hypersetup{
    colorlinks=true,
    linkcolor=blue,
    filecolor=black,      
    urlcolor=blue,
    citecolor=blue,
}
\usepackage{natbib}
\usepackage{braket}
\usepackage{soul}
\usepackage{relsize}
            
\usepackage{pifont}

\newcommand{\Rmnum}[1]{\uppercase\expandafter{\romannumeral #1}}  

\newenvironment{figurehereExtend}
{\def\@captype{figure}

}
{}

\begin{document}

\title{Mechanical non-reciprocity programmed by shear jamming in soft composite solids}
\author{Chang Xu}\thanks{These authors contributed equally}\affiliation{Department of Physics, The Hong Kong University of Science and Technology, Hong Kong SAR, China}
\author{Shuaihu Wang}\thanks{These authors contributed equally}\affiliation{Department of Physics, The Hong Kong University of Science and Technology, Hong Kong SAR, China}
\author{Hong Wang}\thanks{These authors contributed equally}\affiliation{Department of Mechanical Engineering, The Hong Kong University of Science and Technology, Hong Kong SAR, China}
\author{Xu Liu}\affiliation{Department of Mechanical Engineering, The Hong Kong University of Science and Technology, Hong Kong SAR, China}
\author{Zemin Liu}\affiliation{Max Planck Institute for Intelligent Systems, Stuttgart, Germany}
\author{Yiqiu Zhao}\email[]{yiqiuzhao@ust.hk}
\affiliation{Department of Physics, The Hong Kong University of Science and Technology, Hong Kong SAR, China}
\author{Wenqi Hu}\email[]{wenqi@ust.hk}
\affiliation{Department of Mechanical Engineering, The Hong Kong University of Science and Technology, Hong Kong SAR, China}
\author{Qin Xu}\email[]{qinxu@ust.hk}
\affiliation{Department of Physics, The Hong Kong University of Science and Technology, Hong Kong SAR, China}

\begin{abstract}

Mechanical non-reciprocity—manifested as asymmetric responses to opposing mechanical stimuli—has traditionally been achieved through intricate structural nonlinearities in metamaterials. However, continuum solids with inherent non-reciprocal mechanics remain underexplored, despite their promising potential for applications such as wave guiding, robotics, and adaptive materials. Here, we introduce a design principle by employing the shear jamming transition from granular physics to engineer non-reciprocal mechanics in soft composite solids. Through the control of the interplay between inclusion contact networks and matrix elasticity, we achieve tunable, direction-dependent asymmetry in both shear and normal mechanical responses. In addition to static regimes, we demonstrate programmable non-reciprocal dynamics by combining responsive magnetic profiles with the anisotropic characteristics of shear-jammed systems. This strategy enables asymmetric spatiotemporal control over motion transmission, a previously challenging feat in soft materials. Our work establishes a novel paradigm for designing non-reciprocal matter, bridging granular physics with soft material engineering to realize functionalities essential for mechano-intelligent systems.

\end{abstract}
\maketitle

Reciprocity is a fundamental principle in nature,  ensuring the symmetric transfer of physical signals in response to stimuli applied from opposite directions~\cite{maxwell1864calculation, RevModPhys.17.343, charlton1960historical}. Breaking this symmetry through spatiotemporal modulation leads to non-reciprocal dynamics, a phenomenon observed in diverse physical systems such as photonic and phononic circuits~\cite{feng2011nonreciprocal, wang2018observation}, quantum-limited amplifiers~\cite{peano2016topological, malz2018quantum}, and acoustic metamaterials~\cite{PhysRevApplied2019}. In mechanical systems, these non-reciprocal dynamics enable innovative  engineering applications, including enhanced energy harvesting~\cite{Raney2016PNAS,hwang2018input}, improved shock absorption~\cite{wang_nonreciprocal_2023}, and the development of mechanical computing technologies~\cite{yasuda2021mechanical}.

Beyond non-reciprocity in dynamic systems, static non-reciprocity allows for direction-dependent mechanical responses in time-invariant systems~\cite{shaat2020nonreciprocal, dong2024programmable}.  For a static non-reciprocal material, opposite strains ($\pm \gamma$) result in asymmetric shear stresses $\sigma_t(\gamma)\neq \sigma_t(-\gamma)$, or normal stresses $\sigma_n(\gamma) \neq \sigma_n (-\gamma)$, as illustrated in Fig.~\ref{fig:sample}(a). These characteristics are often achieved through the design of mechanical and structural nonlinearities in topological metamaterials. For example, researchers have investigated fishbone structures that violate Maxwell-Betti reciprocity ~\cite{coulais2017static} and  cylindrical lattices that break the symmetry of normal stress responses~\cite{dong2024programmable}. In contrast, mechanical non-reciprocity in continuum solids remains less explored, with Wang {\em et al.} recently reporting a hydrogel-based composite that exhibits an asymmetric response to shear~\cite{wang2023mechanical}. From a materials science perspective, non-reciprocity in continuum media is desirable as it does not depend on system size and geometry. However, our understanding of how to design and program this property remains limited.

In this work, we leverage the concept of shear jamming from granular physics~\cite{bi2011jamming, PAN2023} to establish a generic design framework for soft continuum composites with non-reciprocal mechanics. The non-reciprocity arises from a delicate balance between frictional contact networks and matrix elasticity near the jamming phase boundary, which can be realized and controlled in a broad range of composite systems. By integrating modulated magnetic profiles with shear-jammed states, we further demonstrate programmable non-reciprocal locomotion in these systems, offering a pathway for engineering adaptive materials with direction-dependent mechanics.

\section{Results}

\subsection{Soft composites with shear-jammed inclusions}

\begin{figure*}[t]
\centering
\includegraphics[width=17cm]{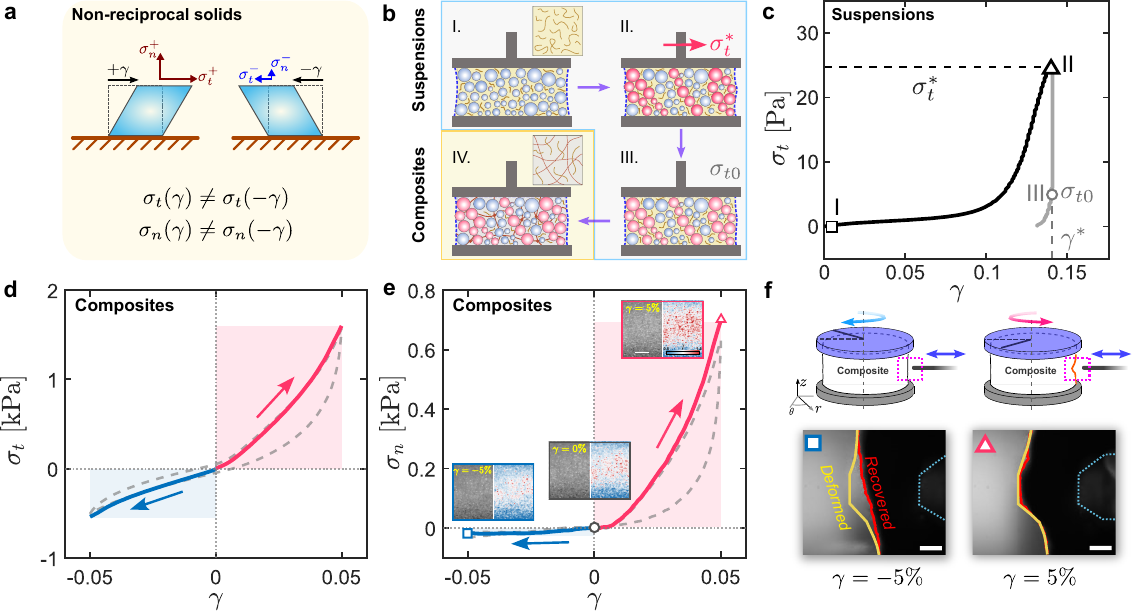}
\caption{\textbf{Soft composites with non-reciprocal mechanics}. {(a) Schematic illustration of materials exhibiting mechanical reciprocity and non-reciprocity. (b) Process of encoding shear-jammed inclusions into soft composites. Steps I to III depict the formation of shear-jammed contact networks during the pre-cured PS–PDMS suspension stages, with the red spheres representing the force-bearing particles. In step IV, the suspension is cured into a composite. (c) Preparation of a jammed network in the pre-cured PS–PDMS suspension. After jamming under a shear strain $\gamma^*$, the shear stress decreases from  $\sigma_t^*$ to $\sigma_{t0}$ while maintaining a constant $\gamma^*$. (d) \& (e) Asymmetric mechanical responses in shear stress ($\sigma_t$) and normal stress ($\sigma_n$) as the composites are sheared from undeformed states, respectively.  Gray dashed loops indicate the corresponding stress-strain hysteresis. Insets in (e) show both unprocessed (gray) and processed (colored) images of the composite-air interfaces under shear. Scale bar: 200~$\mu$m.  (f) Asymmetric shape reversibility.  At $\gamma = 5$~\%, the composite-air interface retains the induced surface deformation, whereas at $\gamma = -5$~\%, the interface fully recovers from surface deformation. Scale bar: 250~$\mu$m.}}
    \label{fig:sample}
\end{figure*}

Soft composites were prepared by first dispersing 22~$\mu$m polystyrene (PS) micro-spheres into a polydimethylsiloxane (PDMS) melt. 
The  fraction of PS spheres was maintained between the shear jamming threshold ($\phi_m=56~\%$) and the static jamming point ($\phi_0 =  69~\%$), ensuring that the PS particles could be jammed by shear~\cite{zhao2024elasticity}. 
With the crosslinkers and catalyst pre-mixed in the PDMS solution, the PS–PDMS suspension was slowly crosslinked into a PS–PDMS composite solid (Figs.~S1 and S2). As illustrated in Fig.~\ref{fig:sample}(b), we implemented a four-step protocol to incorporate a shear-jammed network within the composites. 
First, a PS–PDMS suspension was loaded into a  25~mm parallel-plate shear cell controlled by a rheometer.  An oscillatory pre-shear was applied to prepare the suspension in an initially stress-free, unjammed state. Second, the shear stress ($\sigma_t$) was incrementally increased up to a critical stress $\sigma_{\rm t}^*$, inducing a rigidity transition via shear jamming (Fig.~\ref{fig:sample}(c)). 
Third, shear stress was then reduced from 
$\sigma_{\rm t}^*$ to
$\sigma_{t0} = 5$~Pa
while maintaining 
a constant shear strain,
$\gamma=\gamma^*$~(Fig.~S3). 
Finally, with $\sigma_t = \sigma_{t0} = 5$~Pa, the suspension was cured into a PS-PDMS composite over 12 h. 
A small yet non-zero $\sigma_{t0}$ preserved the jammed particle configurations controlled by $\gamma^*$, while minimizing residual stresses in the resulting composites.
After curing, $\sigma_t$ was reduced to zero, establishing the reference state. 

\begin{figure*}[t]
    \includegraphics[width=16cm]{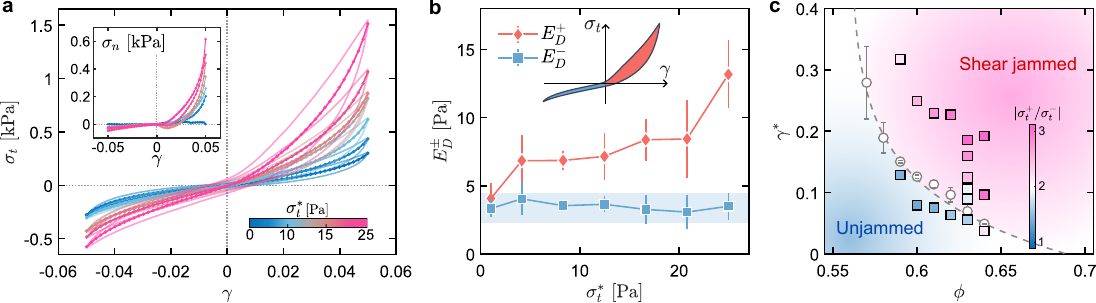}
    \caption{\textbf{Mechanical non-reciprocity controlled by shear jamming transition.} (a) Plots of shear stress ($\sigma_t$) against shear strain ($\gamma$) for PS–PDMS composites with a constant $\phi = 0.63$ prepared under varying critical shear stresses ($\sigma^*_t $) ranging from 1~Pa to $25$~Pa. The connected dots represent the mean values of $\sigma_t(\gamma)$ obtained by averaging hysteresis loops at each strain. Inset: plots of associated non-reciprocal normal stress ($\sigma_n$) against $\gamma$ for these composites.  (b) Plots of hysteresis loop areas $E_D^+$ and $E_D^-$ against $\sigma^*_t$. 
    (c)
    Relationship between the shear jamming transition of inclusion and the degree of shear non-reciprocity of the resulting composites. Open circles represent the boundary of shear-jamming transition for PS–PDMS suspensions, fitted to Eq.~\ref{eqn: SJ} (dashed line). Squares mark the material parameters $(\phi, \gamma^*)$ used to prepare different composite samples, with the filled colors representing the degree of shear non-reciprocity $\vert \sigma_t^+/\sigma_t^-\vert$. The error bars in panels (b) and (c) indicate the standard deviations obtained from the measurements of three to five independently fabricated samples. } 
    \label{fig:program_mech}
\end{figure*}

\subsection{Shear jamming controlled non-reciprocity}
	
Figure~\ref{fig:sample}(d) illustrates the asymmetric shear response of a soft composite with a matrix shear modulus of $G_m=0.25$~kPa and particle volume fraction of $\phi =63~\%$. The PS inclusions were jammed at $\sigma_t^* = 25$~Pa during preparation. Positive shear strain ($\gamma > 0$) is defined as strain in the same direction as the preparation strain $\gamma^*$, while negative strain ($\gamma<0$) corresponds to the opposite direction. At $\gamma = 0.05$, the composite exhibited a shear stress $\sigma_t = 1600$~Pa, approximately three times  the magnitude of the stress $\sigma_t = - 538$~Pa observed at $\gamma = -0.05$. This asymmetric response was highly reproducible under repeated measurements at varying strain amplitudes (Figs.~S6 and S7). Unlike conventional non-reciprocal systems~\cite{wang2023mechanical,coulais2017static,dong2024programmable}, these shear-jammed composites also displayed asymmetric stress-strain hysteresis. As shown by the dashed grey line in Fig.~\ref{fig:sample}(d),  hysteresis is pronounced for $\gamma >0$ but negligible for $\gamma < 0$. 

We attribute these non-reciprocities to the contact configurations encoded in the PS–PDMS suspensions. Previous studies demonstrated that shear-jammed networks are anisotropic, with a major axis aligned at approximately 45$^\circ$ relative to the shear direction~\cite{seto2019shear,Wang2018PRL,zhao2019shear}. Within the composites, anisotropic force networks were preserved by the elastic matrix (Fig.~S9). Consequently, shear along the jamming direction ($\gamma>0$) enhances this anisotropy, increasing shear stress and frictional dissipation~\cite{bi2011jamming,gallier2014rheology}. Conversely, reverse shear ($\gamma<0$) softens the contact networks between inclusions, leaving the matrix as the main contributor to shear stress. The reduction in frictional contacts decreases dissipation under quasi-static shear, resulting in a smaller hysteresis loop for $\gamma<0$.     

The granular characteristics of these soft composites were also evident  in their normal stress responses ($\sigma_n$). As shown in Fig.~\ref{fig:sample}(e),  $\sigma_n$ remains close to zero for $\gamma<0$ but increases sharply for $\gamma>0$, accompanied by a roughening of the composite–air interface (insets in Fig.~\ref{fig:sample}(e) and Supplementary Video 1). Similar to frustrated dilation in shear thickened granular suspensions~\cite{Seto_Giusteri_2018,haitao2024}, the PS particles tended to protrude at composite–air interfaces for $\gamma>0$, while being constrained by the PDMS matrix. This dilation tendency under $\gamma>0$ resulted in pronounced non-reciprocity in $\sigma_n$, a phenomenon referred to as asymmetric Poynting effects in solid mechanics~\cite{poynting1909pressure}. Due to the differences in the governing interaction between $\gamma<0$ and $\gamma>0$, the composites also exhibited asymmetric shape reversibility under deformation (Fig.~\ref{fig:sample}(f)). At $\gamma = -0.05$, the surface deformations induced by an external indentation fully recovered after the indenter was removed. In contrast, at $\gamma = 0.05$, the deformed surface profile persisted indefinitely, even in the absence of external forces. This behavior suggests the presence of a shear-controlled switchable memory for surface deformations (Supplementary Video 2).

Given that the shear-jammed contact networks formed in the suspension state, we investigated the influence of $\sigma_t^*$,  which governs shear jamming of inclusions (Fig.~\ref{fig:sample}), on the non-reciprocity in the resulting PS–PDMS composite. Figure~\ref{fig:program_mech}(a) shows the shear responses of soft composites with constant $G_m=0.25$~kPa and $\phi = 63~\%$, as $\sigma_t^*$ was varied from 1 to 25~Pa. Here, $\sigma_{t}^+$ and $\sigma_{t}^-$ denote the  shear stresses measured at $\gamma = 5~\%$ and $\gamma = -5~\%$, respectively. The degree of shear non-reciprocity, quantified by the ratio $\vert \sigma_t^+/\sigma_t^- \vert$, increases with $\sigma_t^*$, rising from approximately~1.1 at $\sigma_t^* = 1$~Pa to~3.3 at $\sigma_t^* = 25$~Pa. The non-reciprocal normal responses also exhibited a strong dependence on $\sigma_t^*$. As shown in the inset of Fig.~\ref{fig:program_mech}, {$\sigma_n$} for $\gamma > 0$ increases with {$\sigma_t^*$} but remains nearly zero for $\gamma < 0$.  Finally, the stress-strain hysteresis varied with $\sigma_t^*$ as well. Using the loop integrals ($E_{D}^+ = \oint_{\gamma>0} \sigma_t d\gamma$) and ($E_{D}^- = \oint_{\gamma<0} \sigma_t d\gamma$) to characterize dissipations in the positive and negative shear directions, respectively, we observed that $E_D^+$ increased substantially with $\sigma_t^*$ while $E_D^-$ remained nearly constant (Fig.~\ref{fig:program_mech} (b)). This further confirmed that $\sigma_t^*$ controls the contact interactions among PS inclusions for $\gamma>0$.

The shear-induced non-reciprocal behaviors of PS–PDMS composites were quantitatively connected to shear jamming phase transitions in PS inclusions (Fig.~\ref{fig:program_mech}(c)). The phase boundary, represented by the open circles, was described by an empirical formula:
\begin{equation}
    \phi(\gamma_J) = \phi_m + (\phi_0 - \phi_m) e^{-\gamma_J/\gamma_c}
    \label{eqn: SJ}
\end{equation}
with $\phi_0 =0.689 \pm 0.004, \phi_m = 0.560 \pm 0.002$, and a characteristic strain $\gamma_c= 0.104 \pm 0.003$~(Fig.~S4). In addition to mapping the phase boundary, we simultaneously plotted $\vert \sigma_t^+/\sigma_t^- \vert$ of soft composites prepared with varying $\phi$ and $\gamma^*$ in Fig.~\ref{fig:program_mech}(c). Below $\phi(\gamma_J)$, $\vert\sigma_t^+/\sigma_t^-\vert$ approaches unity, indicating a low degree of shear non-reciprocity. In contrast, this ratio exceeds unity above $\phi(\gamma_J)$ and generally increases with the distance to $\phi(\gamma_J)$. 

Besides PS-PDMS composites, we used different particulate inclusions–including glass (Fig.~S5),  poly(methyl methacrylate) (PMMA) (Fig.~S5), and Sodium Chloride (NaCl)–to engineer composites with shear-jammed configurations. Notably,  glass- and PMMA-embedded composites exhibited similar non-reciprocal behaviors to PS–PDMS composites, underscoring the broad applicability of the proposed design principle.
For composites with hexahedral  NaCl particles, the non-reciprocity was enhanced by up to tenfold in stress difference and more than fiftyfold in modulus difference (Extended Data Fig.\ref{fig:Enhanced}), matching the reported performance in Ref.~\cite{wang2023mechanical}.

\begin{figure}
    \centering
    \includegraphics[width=8cm]{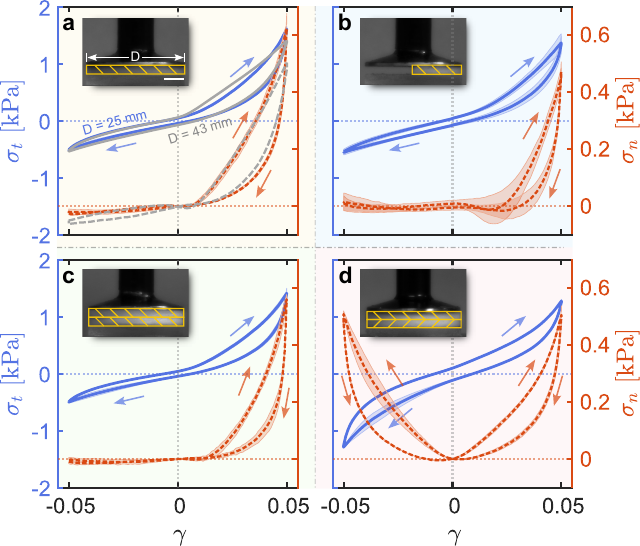}
    \caption{\textbf{Scalable and programmable mechanical non-reciprocity.} (a) Plots of non-reciprocal shear stress ($\sigma_t$) and normal stress ($\sigma_n$) for the composites with the diameters of $D=25$~mm (color) and $D = 43$~mm (grey), respectively. Scale bar: 5 mm. (b) Plots of $\sigma_t$ and $\sigma_n$ against $\gamma$ for half-disk samples cut from the $D=25$ composite shown in (a). To minimize the sample-to-sample variations induced by cutting. Panel (b) shows the results by averaging four individual measurements. (c) Plots of $\sigma_t$ and $\sigma_n$ against $\gamma$ for a two-segment sample created by stacking two composites with force chains aligned in parallel.  (d) Plots of $\sigma_t$ and $\sigma_n$ against $\gamma$ for a two-segment sample formed by stacking two composites with force chains aligned perpendicularly.} 
    \label{fig:prog}
\end{figure}

\subsection {Scalable and programmable mechanical non-reciprocity}

As the bulk of composites is spatially homogeneous~(Fig.~S8), the non-reciprocal responses are size independent. Figure~\ref{fig:prog}(a) demonstrates this for soft composites with identical material parameters ($G_m = 0.25$~kPa, $\phi = 63$~\%, and $\sigma_t^* = 25$~Pa) but different diameters ($D = 25$~mm and $43$~mm), where the asymmetric responses of $\sigma_t(\gamma)$ and $\sigma_n(\gamma)$ remain consistent. Furthermore, cutting the $R=25$~mm composite in half along its diameter yielded a half-disk sample. As shown in Fig.~\ref{fig:prog}(b), both $\sigma_t$ and $\sigma_n$ remain asymmetric against $\gamma$, mirroring the responses of a full composite. These results demonstrate that the observed mechanical non-reciprocity is an inherently scalable and fracture resistant material property.

Furthermore, the shear jamming controlled non-reciprocity enables programmable mechanics in multi-segment composites. As shown in Fig.~\ref{fig:prog}(c), for a two-segment composite with aligned jamming directions, both $\sigma_t$ and $\sigma_n$ exhibit the same asymmetric responses to cyclic shear as a single composite. In contrast, a two-segment composite with opposite jamming configurations (Fig.~\ref{fig:prog}(d)) shows symmetric responses against $\gamma$, accompanied by enhanced hysteresis for $\gamma<0$, suggesting that direct particle contacts dominate mechanical responses in both shear directions.

\begin{figure*}
    \centering
    \includegraphics[width=16cm]{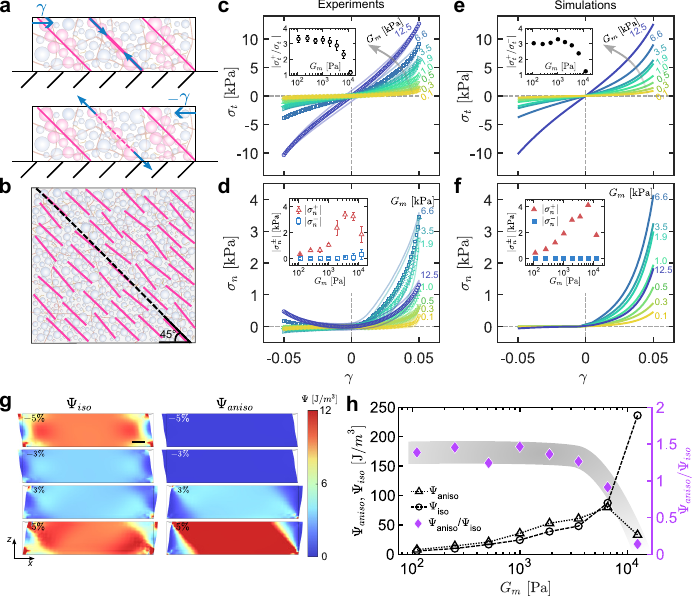}
    \caption{\textbf{Continuum constitutive model.} (a) Modelling of force chains as semi-rigid fibers within composites, which can resist compression (top) but fail under tension (bottom). (b) Schematics of the fiber-reinforced matrix for non-reciprocal composite modelling. Pink lines:  fibers in composites dispersed around a mean orientation, approximately $45^\circ$ to the boundary. (c) \& (d) Experimental results of $\sigma_t$ and $\sigma_n$ against $\gamma$ as $G_m$ ranges from 0.1 to 12.1~kPa, and $\phi = 63~\%$ and $\sigma_t^* = 25$~Pa remain constant. Inset of (c): experimental results of $\vert \sigma^+_t/\sigma^-_t\vert$ versus $G_m$. Inset of (d): experimental results of $\sigma^+_n$ and $\sigma^-_n$ against $G_m$, respectively. The error bars represent the standard deviations from at least three independent measurements. (e) and (f): Model predictions of $\sigma_t$ and $\sigma_n$ against $\gamma$, respectively, for different $G_m$ ranging from 0.1 to 12.1~kPa ($\phi = 63~\%$ and $\sigma_t^* = 25$~Pa). Inset of (e): simulation results of $\vert \sigma^+_t/\sigma^-_t \vert$ versus $G_m$. Inset of (f): simulation results of $\sigma^+_t$ and $\sigma^-_t$ against $G_m$, respectively. (g) Distribution of strain energy density ($\Psi_{\rm iso}$ and $\Psi_{\rm aniso}$) inside a composite ($G_m = 0.25$~kPa, $\phi=63~\%$ and $\sigma_t^* = 25$~Pa) under cyclic shear, derived from the fiber-reinforced model. Scale bar: $2$ mm. (h) Left y-axis: plots of $\Psi_{\rm iso}$ and $\Psi_{\rm aniso}$ against $G_m$. Right y-axis: plot of $\Psi_{\rm aniso}/\Psi_{\rm iso}$ against $G_m$.}
    \label{fig:Matrix}
\end{figure*}

\subsection{Continuum constitutive model}

To elucidate the observed non-reciprocal responses, we develop a continuum constitutive model that treats the composites as fiber-reinforced anisotropic materials~\cite{Spencer1986}. 
The anisotropic force chains arising from shear jamming are represented as a network of semi-rigid fibers that resist compressive loads but exhibit vanishing rigidity under tension (Fig.~\ref{fig:Matrix}(a)). This framework leads to a strain energy density function, decomposed as $\Psi =\Psi_{\rm vol}+\Psi_{\rm iso}+\Psi_{\rm aniso}$, where $\Psi_{\rm vol}+\Psi_{\rm iso}$ accounts for the hyperelastic response of the deformed matrix, while $\Psi_{\rm aniso}$ captures the directional reinforcement contributed by the embedded fibers (Fig.~\ref{fig:Matrix}(b)). 

Given a bulk modulus $K$ and a deformation gradient $\mathbf{F}=\partial x/\partial X$, the volumetric strain energy density is
\begin{equation}
    \Psi_{\rm vol} = \frac{K}{2} \left(J-1\right)^2,
    \label{eqn:vol}
\end{equation} with $ J = \det\left( \mathbf{F} \right)$. 
To account for the nonlinear elasticity of the soft silicone matrix under volume-conserving deformations, the  isochoric component $\Psi_{\rm iso}$ is characterized by a two-term Yeoh model \cite{yeoh1993some},
 \begin{equation}
     \Psi_{\rm iso} = \frac{\mu_1}{2}\left(\bar{I}_1 - 3\right) + \frac{\mu_2}{2} (\bar{I}_1 - 3)^2.
     \label{eqn:iso}
\end{equation}
Here, $\mu_1$ denotes the baseline shear modulus, $\mu_2$ is a material parameter governs nonlinear stiffening, and $\bar{I}_1=J^{-2/3} I_1$ is the modified first invariant of the Cauthy--Green strain tensor $\mathbf{C} = \mathbf{F}^T\cdot \mathbf{F}$, with $I_1= Tr(\mathbf{C})$. 
The anisotropic contribution from shear-jammed force chains is modeled through a modified Holzapfel-Gasser-Ogden (HGO) formulation~\cite{holzapfel2002nonlinear} 
\begin{equation}
    \Psi_{\rm ainso}= \frac{k_1}{2\beta} \left\{ e^{\beta\left[ \kappa \bar{I}_1 +(1-3\kappa)\bar{I}_4-1 \right]^2} -1 \right\} H(1-\lambda_a), 
    \label{eqn:aniso}
\end{equation}
where $k_1$ (fiber stiffness), $\kappa$ (fiber dispersion), and $\beta$ (dimensionless nonlinearity parameter) characterize the composite’s directional reinforcement. 
The pseudo-invariant $\bar{I}_4 =J^{-2/3} I_4$ incorporates the directional stretch $I_4=\mathbf{a}\cdot \mathbf{Ca}$, with $\mathbf{a}= (-1, 0, 1)/\sqrt{2} $ representing the mean fiber orientation. Critically, the Heaviside function $H(1-\lambda_a)$ enforces the fibers’ semi-rigid response, deactivating their contribution under tension  ($\lambda_a = \vert \vert \mathbf{Fa} \vert\vert \geq 1$). 

We directly compared the experimentally measured non-reciprocal $\sigma_t$ and $\sigma_n$ with predictions from our constitutive model. To probe the interplay between matrix elasticity and frictional contacts, we varied the matrix shear modulus $G_m$ from 0.11~kPa to 12.5~kPa, while maintaining a constant $\phi = 0.63$ and $\sigma_t^* = 25$~Pa. Figure~\ref{fig:Matrix}(c) shows the resulting $\sigma_t(\gamma)$ and $\sigma_n (\gamma)$ for composites across this range. For $G_m \leq 2$~kPa, the shear non-reciprocity ratio $\vert \sigma_t^+/\sigma_t^- \vert$ remains nearly constant at~$\sim 3.3$, but sharply declines to 1.4 for $G_m >2$~kPa (Fig.~\ref{fig:Matrix}c, inset), indicating that excessive matrix stiffness reduces shear-jammed contact networks. Additionally, the non-reciprocal normal responses exhibit a distinct dependence on $G_m$. Analogous to the shear response, we define $\sigma^+_n$ and $\sigma^-_n$ as the normal stresses measured at $\gamma = 5$\% and $-5$\%, respectively. While $\vert \sigma^-_n \vert$ remained insignificant across all tested $G_m$, $\vert \sigma^+_n \vert$ was substantially larger than $\vert \sigma^-_n \vert$ and demonstrated a strong dependence on $G_m$. Notably, $\vert \sigma^+_n \vert$ increased from 0.36~kPa to 3.58~kPa  as $G_m$ rose from 0.1~kPa to 6.6~kPa, before decreasing to 1.46~kPa at $G_m$ =12.5~kPa. The non-monotonic dependence of $\sigma_n$ on $G_m$ suggests that matrices with low crosslinking densities stabilize frictional force chains, whereas highly crosslinked matrices suppress the strength of force chains.

Finite element method (FEM) simulations (see Method and Table S1) show that the continuum constitutive model (Eqs.~\ref{eqn:vol} to \ref{eqn:aniso}) effectively predicts non-reciprocal $\sigma_t$ and $\sigma_n$ for different matrix elasticity ($G_m$) (Figs.~\ref{fig:Matrix}(e) and (f)), aligning with the experimental measurements. 
Figure~\ref{fig:Matrix}(g) presents the simulated isotropic ($\Psi_{\rm iso}$) and anisotropic ($\Psi_{\rm aniso}$) energy densities under cyclic shear for a composite with $G_m = 0.25$~kPa. While $\Psi_{\rm iso}$ remains symmetric with respect to $\gamma$,  $\Psi_{\rm aniso}$ exhibits pronounced asymmetry between $\gamma>0$ and $\gamma<0$, directly leading to non-reciprocal mechanics of composites. To quantify the competition between matrix and force chains, we analyzed $\Psi_{\rm iso}$ and $\Psi_{\rm aniso}$ at $\gamma=5$~\% across varying $G_m$ in Fig.~\ref{fig:Matrix}(h). For $G_m \leq 2$~kPa, the ratio $\Psi_{\rm aniso}/\Psi_{\rm iso}$ remains constant and exceeds unity, confirming the dominance of force chains.  Beyond this threshold ($G_m > 2$~kPa), the ratio declines sharply, reflecting the suppression of anisotropic networks in overly rigid matrices. These experimental and computational results collectively confirm that non-reciprocal mechanics emerge from a delicate balance between sufficient matrix compliance to preserve shear-jammed anisotropy and excessive rigidity that disrupts force chain strength.

\begin{figure*}[t]
    \centering
    \includegraphics[width=16cm]{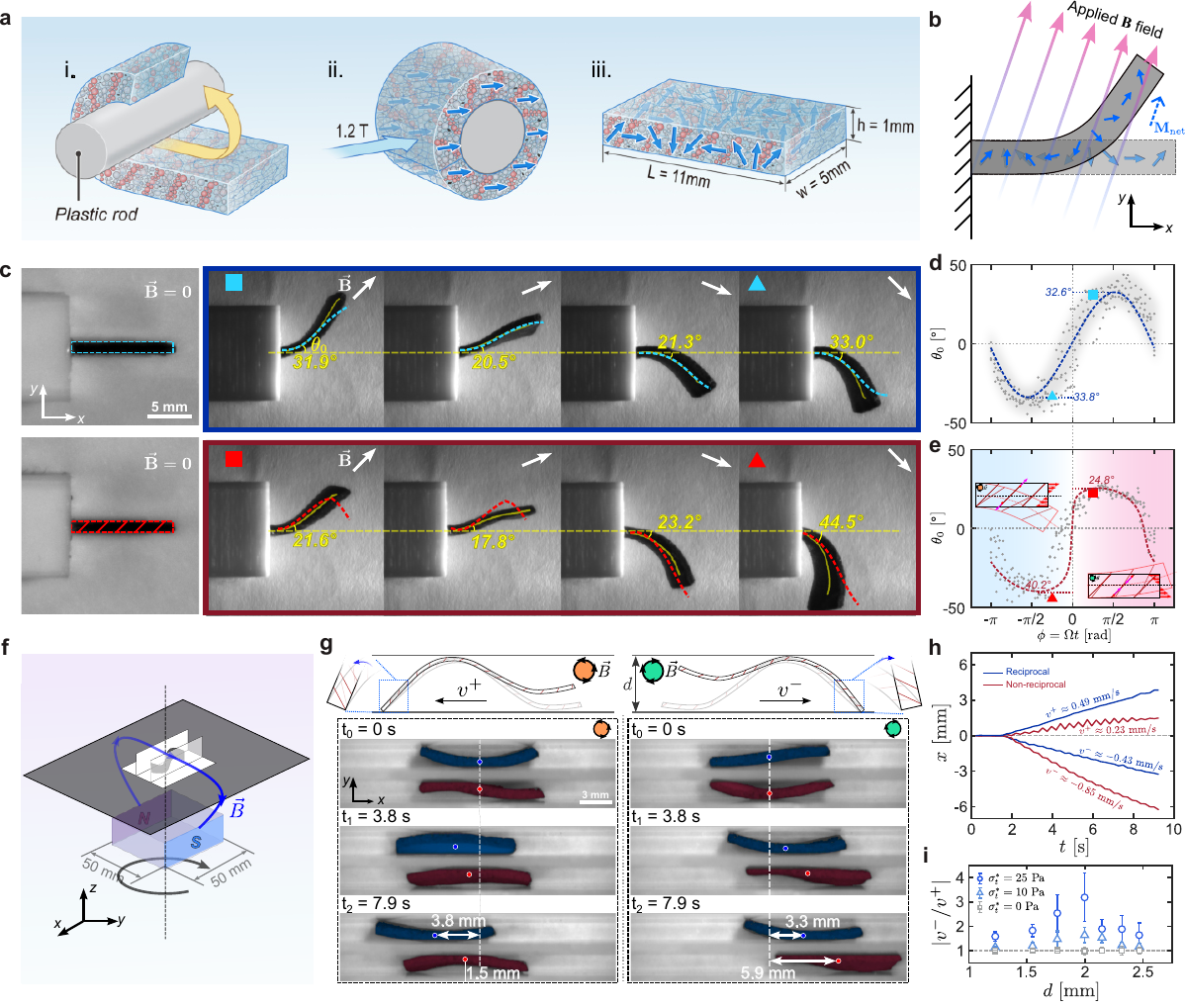}
    \caption{\textbf{Non-reciprocal active solids}.  (a) Fabrication of soft composites incorporating shear-jammed PS inclusions and spatially-modulated magnetic domains.  (b) Schematic illustration of a composite sheet bending under a rotational magnetic field $\mathbf{B}$. (c) Snapshots of an isotropic composite (top: without shear-jammed inclusions) and an anisotropic composite (bottom: with shear-jammed inclusions) bending under a rotational field with $B_0 = 50$~mT and $\Omega = 0.23$~rad/s. Solid yellow lines represent the centerlines of the bending composites, while the colored dashed lines indicate the numerical predictions derived from Eq.~\ref{eqn:torque_balance}. (d) Plot of the deflection angle ($\theta_0$) against the orientation of the rotating magnetic field ($\phi = \Omega t$) for the isotropic composite. Gray points represent experimental results and the blue dashed line represents the numerical prediction derived from Eq.~\ref{eqn:torque_balance}. (e) Plot of $\theta_0$ against $\phi$ for the anisotropic composite. Gray points indicate experimental results and the red dashed line shows the numerical prediction obtained by combining the fiber-reinforced model with Eq.~\ref{eqn:torque_balance}. Inset: schematic illustration of fibers under compression or tension due to bending. (f) Schematic of the experimental setup where a rotating magnetic ($\mathbf{B}$) drives the locomotion of sheet-shaped composites in a confined space with gap size $d$. (g) Depending on the rotational direction of the $\mathbf{B}$ field, a composite can move in either the $(+)$ direction or $(-)$ direction. An isotropic composite (blue) can move in both directions with approximately symmetric speeds ($v^-\approx v^+$), while an anisotropic composite (red) moves substantially faster in the $(-)$ direction than in the $(+)$ direction ($v^- >v^+$). (h) Plots of the horizontal displacement ($x$) versus time ($t$) for both reciprocal and non-reciprocal composites. (i) Plot of the ratio $\vert v^-/v^+ \vert$ against the gap size $d$.  The error bars indicate the standard deviations obtained from the measurements of three independently fabricated samples.}
    \label{fig:activesolid}
\end{figure*}

\subsection{Non-reciprocal active solids}

Beyond static mechanical non-reciprocity, we extended our soft composite design strategy to engineer active solids with non-reciprocal dynamics. 
To this end, we first prepared a mixture of PS particles and NdFeB beads (radius: 5~$\mu$m) in a PDMS solution, with volume fractions of $\phi_{\rm PS} = 62~\%$ and $\phi_{\rm NdFeB} = 1$~\%, respectively. 
The mixture was shear-jammed under a critical stress $\sigma_t^* = 25$~Pa and subsequently cured into a composite with $G_m =1$~kPa. To program activity, a section of the composite was rolled around a plastic rod (radius: $1.75$~mm) and exposed to a $1.2$~T magnetic field (Fig.~\ref{fig:activesolid}(a)). This process encoded a modulated magnetization profile along the material coordinate $s$, described by:  
\begin{equation}
\mathbf{m} = m_0 [\cos{(\omega_s s+\psi_0)}, \sin (\omega_s s +\psi_0), 0],
\end{equation} 
where $m_0$ is the magnetization magnitude, $\omega_s$ defines the spatial frequency, and $\psi$ represents the initial phase of the magnetization profile. 

As shown in Fig.~\ref{fig:activesolid}(b), the composite sheet exhibits global bending under an external magnetic field due to its non-zero local magnetization. When subjected to a rotating magnetic field $\mathbf{B} = B_0 (\cos{\Omega t}, \sin{\Omega t}, 0)$ with $B_0 = 50$~mT, the sheet oscillates in the $x-y$ plane with the same angular frequency $\Omega$. Snapshots in Fig.~\ref{fig:activesolid}(c) reveal that,  for a PS-PDMS composite with unjammed inclusions ($\sigma_t^*=0$~Pa),  symmetrical bending occurs as $\mathbf{B} $ rotates clockwise with $\Omega =0.23$~rad/s (top panels). In contrast,  a composite with shear jammed inclusions ($\sigma_t^*=25$~Pa) displays marked bending asymmetry, deflecting downward by a maximum of 44.5$^\circ$ but upward by only 21.6$^\circ$  (bottom panels).  

For a composite with a cross-sectional area $A$, the bending profile results from the local torque balance along the $z$–direction:
\begin{equation}
    -\frac{\partial M_b}{\partial s} = \tau_m A = [ A \mathbf{R_z (\theta)} (\mathbf{m} \times \mathbf{B})] \vert_z,
    \label{eqn:torque_balance}
\end{equation}
where $\mathbf{R_z (\theta)}$ is the rotational matrix in the $x–y$ plane~\cite{Lum2016shape,hu2018small}. For reciprocal isotropic composites, $M_b = E I \partial \theta/\partial s$, where $E$ is Young's modulus and $I$ is the area moment of inertia. By solving for $\theta(s)$ using Eq.~\ref{eqn:torque_balance}, we derived the profiles of a reciprocal composite sheet under a rotating magnetic field $\mathbf{B}$, as shown by the blue dashed line in Fig.~\ref{fig:activesolid}(c). Additionally,  the dashed line in Fig.~\ref{fig:activesolid}(d) plots the predicted deflection angle $\theta_0$ (defined by $\theta \vert_{s=0}$) against the orientation of $\mathbf{B}$, $\phi = \Omega t$, demonstrating consistency with the experimental results.

\begin{figure*}
\centering
\begin{figurehereExtend}
    \centering
    \includegraphics[width=15cm]{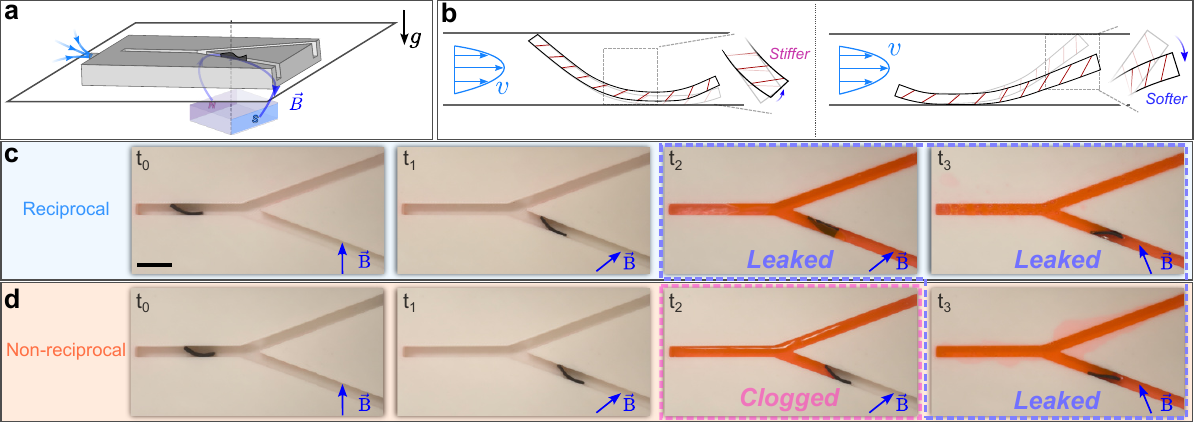}
    \caption{\textbf{Selective flow control with non-reciprocal active solids}. (a) Schematic of the experimental setup where a sheet-shaped composite is placed in a “Y” chamber (inner diameter: 3.3~mm). A movable rotational magnet, attached beneath the chamber, controls the movement and shape change of composite samples. The magnitude of the magnetic field was held constant at $B=15$ mT. The Rhodamine-dyed water is then injected into the left inlet of the Y-chamber. (b) Schematic illustrations of the asymmetric bending for non-reciprocal composites within a flow channel. The left and right panels highlight the bending differences resulting from anisotropic shear-jammed networks. (c) A reciprocal composite fails to block the channel flow in either anchoring directions. The snapshots at four time points ($t_0$, $t_1$, $t_2$, and $t_3$) are presented to demonstrate this response.  $t_0$: Initial state. $t_1$: The sample migrates to the lower branch due to the movement of an external rotating magnet field. $t_2, t_3$: Water leaks occur in both anchoring configurations. (d) The non-reciprocal composite achieves switchable blocking and releasing of the channel flow. At $t_2$, the water flow is blocked. By reversing the bending configuration at $t_3$, the water leak occurs again. Scale bar: $1$ cm. The blue arrow represents the direction of the external $\mathbf{B}$ field direction.}
    \label{fig:clogged}
\end{figurehereExtend}
\end{figure*}

For non-reciprocal composites, however,  the elastic torque $\partial M_b / \partial s$ can not be evaluated analytically due to the absence of a direct relationship between $M_b$ and $E$. To address this, we re-expressed $\partial M_b / \partial s$ as $\theta^\prime (s)\partial M_b / \partial \theta$~(see the Method), which was then numerically evaluated using the fiber-reinforced anisotropic model (Eqs.~\ref{eqn:vol}–\ref{eqn:aniso}). The red dashed lines in Fig.~\ref{fig:activesolid}(c) represents the theoretically calculated shapes of the non-reciprocal composite, employing the same material parameters as in Fig.~\ref{fig:Matrix}. A direct comparison between model predictions and experimental measurements is further illustrated by the asymmetric $\theta_0(\phi)$ curves in Fig.~\ref{fig:activesolid}(e). The fiber-reinforced model also offers a physical interpretation of the observed bending response. Considering a composite with fibers aligned from the top-right to bottom-left (see the insets in Fig.~\ref{fig:activesolid}(e)), downward bending stretches the fibers, while upward bending compresses them. Due to the inherent asymmetry in the fiber response to stretching versus compression (Eq.~\ref{eqn:aniso}), the composite exhibits bending asymmetry.  
For practical applications, we demonstrate that this bending asymmetry can enable selective flow control in confined channels (Extended Data Fig.~\ref{fig:clogged} and Supplementary Video 5).

Researchers have explored how soft magnetic materials can mimic biological organisms by exhibiting adaptive locomotion in confined spaces~\cite{ziyu2021}. 
Here, we demonstrate that encoded shear-jammed structures enable non-reciprocal locomotion in response to confinement. As illustrated in Fig.~\ref{fig:activesolid}(f), an untethered composite sheet is placed between two acrylic plates with a gap distance $d<3$~mm.  Depending on the rotational direction of the magnetic field ($\mathbf{B}$), the sheet moves either forward $(+)$ or backward $(-)$.  The positive $(+)$ axis corresponds to the direction of pre-shear applied during the shear jamming of the suspensions, while the negative $(-)$ axis denotes the opposite direction. 

\begin{figure*}
\centering
\begin{figurehereExtend}
    \centering
    \includegraphics[width=16cm]{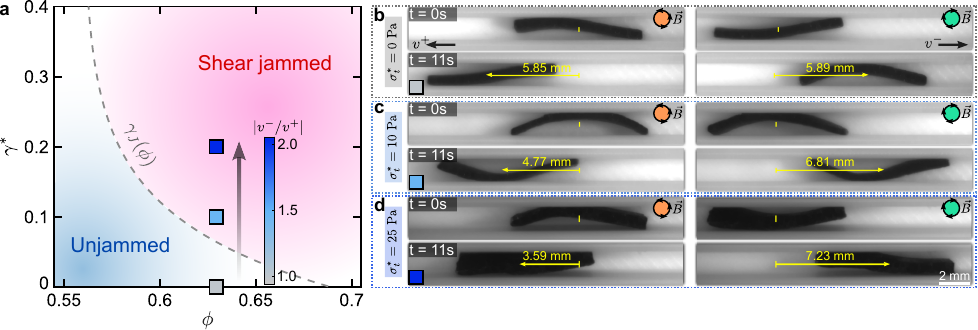}
    \caption{{\bf  Non-reciprocal dynamics programmed by shear-jamming transition.} 
    (a) Relationship between the shear jamming transition of PS particles and the velocity asymmetry of PS-PDMS composites $\vert v^-/v^+\vert$ ($\phi = 63\%$ and $G_m = 1$~kPa). (b–d) Time-lapse snapshots (at $t=0$~s and 11~s) of PS-PDMS composites prepared under different shear stresses ($\sigma_t^* = 0$~Pa, 10~Pa, and 25~Pa) moving in opposite directions within a confined channel  ($d=1.73$~mm). }
    \label{fig:PE}
\end{figurehereExtend}
\end{figure*}

We prepared two sheet-shaped samples: an isotropic composite fabricated without pre-shear and an anisotropic composite with shear jammed inclusions formed under $\sigma_t^ * = 25$~Pa. 
Both composites shared identical matrix elasticity ($G_m= 1$~kPa) and ($\phi =63~\%$). Snapshots in Fig.~\ref{fig:activesolid}(g) directly compare their locomotion dynamics under confinement (gap size $d=2$~mm) when exposed to a rotating magnetic field ($\mathbf{B}$) with $B_0 = 50$~mT and $\Omega = 14.3$~rad/s. The isotropic composite (blue)  exhibited symmetric motion, displacing approximately $\sim$3.5~mm in both directions over 7.9~s. In contrast, the anisotropic composite (red) demonstrated non-reciprocal dynamics, displacing $5.8$~mm in the ($-$) direction but only $1.6$~mm in the ($+$) direction over the same period (Supplementary Video 3). 
Figure~\ref{fig:activesolid}(h) plots the horizontal displacements $x(t)$ for both composites, showing zig-zag oscillations consistent with a frequency of $\Omega= 14.3$~rad/s. We defined directional velocities $v^+ = x^\prime (t)\vert_{x>0}$ and  $v^-= x^\prime (t)\vert_{x<0}$  from the mean slopes of these curves. For the anisotropic composite, $v^- \approx  -0.85$~mm/s approximately three times $v^+ \approx 0.23$~mm/s, highlighting its directional bias. The degree of dynamic non-reciprocity, quantified by $\vert v^{-}/v^{+} \vert$, show non-monotonic dependence on the gap size $d$: increasing $d$  from $1.2$~mm to $2.5$~mm, $\vert v^{-}/v^{+} \vert$ first increases, peaked at $\sim 3.2$ at $d \approx 2$~mm, then decreases (Fig.~S11 and Supplementary Video 4). This non-reciprocal dynamics can be quantitively tuned based on the shear jamming phase diagram (Extended Data Fig.~\ref{fig:PE}). Figure~\ref{fig:activesolid}(i) presents the plots of $\vert v^{-}/v^{+} \vert$ against $d$ for $\sigma^*_t=0$~Pa, 10~Pa, and 25~Pa, respectively.

\begin{figure*}
\centering
\begin{figurehereExtend}
    \centering
    \includegraphics[width=15.0cm]{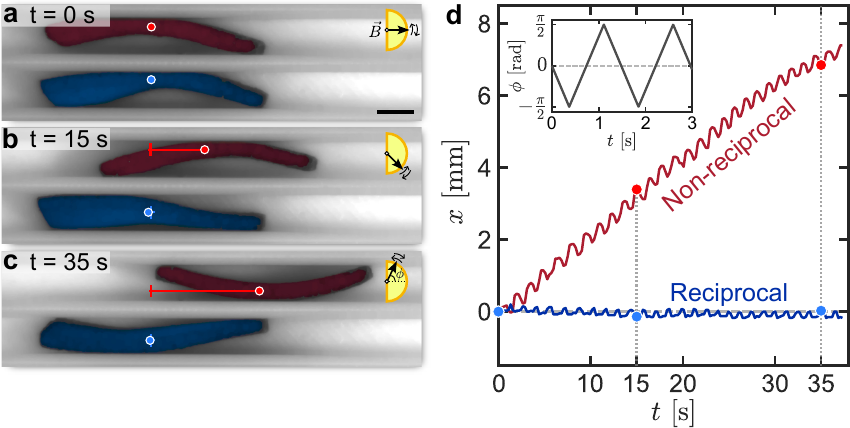}
    \caption{\textbf{Unidirectional motion of non-reciprocal composites under an oscillating magnetic field}. (a – c) Time-lapse snapshots (at $t=0$~s, $15$~s, and $35$~s) for both reciprocal ($\sigma_t^*= 0$~Pa, blue) and non-reciprocal composites ($\sigma_t^*=25~$Pa, red)  under an oscillating magnetic field.  The field direction oscillates between $\phi = -\pi/2$ and $\phi = \pi/2$ at a frequency of $1.1$~Hz while the field magnitude remains constant at $B=50$~mT.   Both composites have the same particle volume fraction ($\phi=63~\%$) and matrix modulus ($G_m = 1~$kPa). (d) Plots of horizontal displacement ($x$) against time ($t$) for both composites.  The reciprocal sample locally oscillates without net movement, whereas the non-reciprocal sample breaks the time-reversal symmetry to achieve directed motion.  Inset: Orientation angle ($\phi$) of the rotating field. }
    \label{fig:OscillatingB}
\end{figurehereExtend}
\end{figure*}

This directional dynamics of the anisotropic sample arises from its asymmetry bending compliance. For the composite in Fig.~\ref{fig:activesolid}(g), motion depended on anchoring its leading edge to the bottom plate during magnetic field rotation, followed by buckling deformation. Upon release of the leading edge from the bottom plate, the composite advanced stepwise within the confined space~\cite{ziyu2021}. The step frequency matched the magnetic field’s rotational frequency ($\Omega$). Anisotropic force chains (Fig.~\ref{fig:activesolid}(g)) introduced asymmetric buckling: elongation of force chains occurred during motion in the $(-)$ direction, while compression dominated in the $(+)$ direction. This asymmetry amplified bending in the $(-)$ direction, resulting in faster motion ($v^->v^+$).  Under strong confinement ($d<1.5$~mm),  buckling suppression in both directions restored near-symmetric motion ($v_-  \approx v_+$). Conversely, in weak confinement ($d\approx 2.5$~mm), unstable anchoring of the leading edge reduced the speed asymmetry ratio $\vert v_-/v_+ \vert$, as shown in Fig.~\ref{fig:activesolid}(i). 
By further exploiting this bending asymmetry through an oscillating magnetic field, we can achieve complete unidirectional motion through time-reversal symmetry breaking (Extended Data Fig.~\ref{fig:OscillatingB} and Supplementary Video 6).

\section{Discussion and Conclusion}

This study introduces a novel paradigm for engineering non-reciprocal soft continuum solids by encoding shear-jammed contact networks during the pre-cured state. By leveraging the intrinsic relationship between shear jamming transitions and the governing interactions in soft composites, we demonstrate how static non-reciprocity – in both shear and normal responses–can be quantitatively tuned by controlling the proximity to shear-jamming boundary ($\phi(\gamma^*)$) and matrix elasticity ($G_m$). This mechanics is robustly captured by a continuum constitutive model unifying nonlinear elasticity with semi-rigid fiber mechanics, which elucidates the critical interplay between contact networks and matrix. In addition to static non-reciprocity, we further show that the shear-jammed configurations synergize with spatially-modulated magnetic domains to achieve programmable non-reciprocal dynamics, enabling a pathway for asymmetric actuation in soft robotic systems.

These findings establish a shear-jamming-based framework for designing soft materials with precisely tailored non-reciprocal functionalities. Given the prevalence of shear jamming in diverse particulate systems, we anticipate that further enhancement of non-reciprocal behavior can be achieved by engineering particle shapes and surface roughness. Beyond mechanical non-reciprocity, our approach could enable asymmetric electrical or thermal transport through the incorporation of jammed conductive particles (Fig.~S14), opening new avenues for energy-efficient material design.

\vspace{11 pt}

\noindent{ \bf METHODS}

\vspace{5 pt}
{\em Material preparation} --- The Polydimethylsiloxane (PDMS) gel matrix was prepared by mixing a silicone base polymer (Gelest, DMS-V31) with a cross-linker copolymer (Gelest, HMS-301) and a Pt-based siloxane complex as catalyst (Gelest, SIP6831.2) \cite{zhao2022the}.  The catalyst concentration was maintained constant at $0.0061\%$, such that the mixture remained  liquid during the first hour and then fully cured within 12 h~(Fig.~S1). This slow curing process enabled the formation of shear-jammed states in the suspension states. The radius of polystyrene (PS) particles followed a logarithmic normal function $f(r)=\exp\left[ - (\ln(r/r_a))^2/{2\sigma^2}  \right]/(\sqrt{2\pi}\sigma r)$ with $r_a = 9.85$~$\mu$m and $\sigma = 0.53$, giving an average  $\left<r\right> = \exp{(\ln{r_a} + \sigma^2/2)} = 11.34~\mu$m (Fig.~S8). The density of PS particles, measured via sedimentation, is $\rho = 1.047\pm0.002~$g/ml. 

\vspace{5 pt}
{\em Composite fabrication} --- Uncured PS-PDMS suspensions were loaded onto a 25~mm parallel-plate shear cell (Anton Paar MCR302 rheometer). Initially, an oscillatory shear, $\gamma(t) = \delta \gamma \sin(\omega t)$ with $\delta \gamma = 10~\%$ and $\omega = 10$~rad/s, was applied for 20~mins to eliminate residual stresses within the mixed suspensions. The resulting smooth surfaces of the relaxed suspensions confirmed a homogeneous, unjammed state after preshear. Next, shear stress gradually increased at a rate of $\dot\sigma_t = 0.1$~Pa/s. Beyond a critical strain $\gamma_J$, the stress-strain slope $d\sigma_t/d\gamma$ rose sharply, signaling the onset of a shear-jamming transition. To mechanically stabilize the jammed network, we further sheared the suspensions to $\sigma^*_t$ under a strain $\gamma^* > \gamma_J$. Subsequently, we reduced $\sigma_t$ to  $\sigma_{t0} = 5$~Pa while maintaining the strain at $\gamma=\gamma^*$ (Fig.~S3), which minimized the applied shear stress during the subsequent curing process. Finally, the jammed suspensions were cured into PS-PDMS composites over 12 hours at $\sigma_t = \sigma_{t0} = 5$~Pa. By varying the weight ratio of crosslinkers from $k = 0.71$~\% to 1.43~\%, the shear modulus of the gel matrix was systematically tuned from $G_m =$ 0.11~kPa to 12.5~kPa.

The active soft composites were prepared using a linear shear device. We found that composites prepared with linear-shear plate and rotational-shear plate show quantitatively similar mechanical non-reciprocity (Figs.~S12 and S13).  Due to the low volume fraction NdFeB beads ($\phi_{\rm NdFeB} = 1$~\%), dipole-dipole interactions within composites are negligible. As a result, the fabricated samples remain flat when freely standing under zero magnetic field.

\vspace{0.2in}

{\em FEM simulation} ---The FEM simulations of the simple shear (Figs.~\ref{fig:Matrix}) were conducted using the commercial software COMSOL Multiphysics (COMSOL Inc., Burlington, MA) with the Structural Mechanics Module (Solid Mechanics Interface). The anisotropic constitutive model was implemented via a user-defined strain energy density determined by  Eqs.~\ref{eqn:vol}–\ref{eqn:aniso}. For the simple shear configuration, a three-dimensional (3D) rectangular domain (20~mm$\times$10~mm$\times$5~mm) was discretized using a swept mesh of hexahedral elements, with a maximum element size of 0.7~mm.  Fixed constraints were applied to the bottom boundary, while the top boundary was permitted only horizontal displacement. We applied a horizontal displacement  in the range from -0.25~mm to 0.25~mm to the top plate, resulting in a shear strain between -5~\% and 5~\%.  For a deformation gradient $\mathbf{F}$,  the stress-strain relation of composites is expressed as
\begin{equation}
\sigma_{ij}=A_{ijkl}\epsilon_{kl}, 
\label{eqn:cons}
\end{equation}
with the fourth-order elastic tensor being defined as
\begin{equation}
 A_{i j k l}=F_{i \alpha} F_{k \beta} \frac{\partial^{2} \Psi}{\partial F_{j \alpha} \partial F_{l \beta}}.
 \label{eqn:tensor}
\end{equation}
Based on Eqs~\ref{eqn:cons} and~\ref{eqn:tensor}, we calculated the shear stress as $\sigma_t = \sigma_{zx}$ (Fig.~\ref{fig:Matrix}(e)) and the normal stress as $\sigma_n=\sigma_{zz}$ (Fig.~\ref{fig:Matrix}(f)). 

The FEM simulations can not effectively reproduce the stress-strain hysteresis caused by frictional interactions among particles. Unlike conventional viscoelasticity, the hysteresis loops of these composites are rate-independent, which can be captured phenomenologically using a Bouc-Wen-based model (Fig.~S10). However, incorporating rate-independent hysteresis into a continuum constitutive relation remains challenging.

\vspace{0.2in}

{\it Material parameters in the constitutive model} ---  To determine the optimal material parameters, we first identified the elastic pre-factors $\mu_1$ and $\mu_2$ by fitting the simulated to the measured $\sigma_t$ for $\gamma<0$, a regime dominated by matrix elasticity.  Subsequently, the fiber stiffness $k_1$ and nonlinear parameter $\beta$ were determined by fitting the simulated to measured $\sigma_t$ for $\gamma>0$, a regime dominated by the shear-jammed inclusions. Given the nearly incompressible nature of  PDMS gels, the bulk modulus was set as $K=100\mu_1$. 
Due to the large value of $K$, the determinant of the deformation matrix ($ J = \det\left( \mathbf{F} \right)$) was consistently close to one, indicating that the dilatational energy density ($\Psi_{vol}$) was negligible compared to the volume-conserving terms ($\Psi_{iso}$ and $\Psi_{aniso}$).
To approximate the slight alignment dispersion of such systems in the FEM simulations, we maintained the fiber parameter as $\kappa = 1/12$~\cite{hou2020shear}, consistent with structure of shear-jammed force networks within composites (Fig.~S9). Within the selected range of material parameters, the simulation results appeared to be insensitive to $K$ and $\kappa$. The detailed material parameters are presented in Supplementary Table 1. 

\vspace{0.2in}

\vspace{0.2in}

{\em Numerical analysis of bending profiles} --- Considering a local deflection angle $\theta$ in the bended composite, the rotational matrix $\mathbf{R_z (\theta)}$ is given by
\[\mathbf{R_z}(\theta) = \begin{bmatrix}
\cos\theta & -\sin\theta & 0 \\
\sin\theta & \cos\theta & 0 \\
0 & 0 & 1
\end{bmatrix}. 
\]
Consequently, Eq.~\ref{eqn:torque_balance} can be explicitly expressed as
\begin{equation}
\begin{aligned}
    &{\partial M_b}/{ \partial s} = -AB_0m_0\\
    & \left\{ [\sin(\Omega t)\cos(\omega_s s+\psi_0)  - \right.  \cos(\Omega t)\sin(\omega_ss +\psi_0)]\cos(\theta)  \\
      & - [\sin(\Omega t)\sin(\omega_s s+\psi_0) + 
    \left. \cos(\Omega t)\cos(\omega_ss +\psi_0)]\sin(\theta)\right\}.
\end{aligned}
    \label{eq:NRBending}
\end{equation}
For the reciprocal composites, we have $\partial M_b /\partial s= E I \partial^2 \theta/\partial s^2$. Thus, a direct integration of Eq.~\ref{eq:NRBending} yielded $\theta(s)$, which determines the bending profiles of reciprocal composites. By comparing the experimental profile and theoretically obtained $\theta(s)$ at $t=0$, we fitted the optimal $\phi_0$ and $C_1 = - A B_0 m_0 / EI$. Bending profiles at any given time $t>0$ were subsequently calculated as shown in Figs.~\ref{fig:activesolid}(c) and (d).

For the non-reciprocal composites, we considered $\partial M_b / \partial s = \theta^\prime (s)\partial M_b / \partial \theta$, where the term $\partial M_b /\partial \theta$ was calculated using FEM simulations. To align with the experimental geometry in Fig.~\ref{fig:activesolid}, a computational unit domain (1~mm$\times$5~mm$\times$1~mm) with a swept mesh grid (maximum element size: 0.275~mm) was defined. While the left boundary was fixed, we applied a vertical displacement ranging from $-0.33$~mm to $0.33$~mm to the right boundary, and measured the bending momentum $M_b$ for different deflection angles $\theta$. Using the simulated $\partial M_b/\partial \theta$ along with the fitted $\phi_0$ and $C_2=(- A B_0 m_0)$, we further obtained $\theta(s)$ and bending profiles of the non-reciprocal composites, as shown in  Figs.~\ref{fig:activesolid}(c) and (e).

\vspace{11pt}
\noindent{\bf  Acknowledgments}

\noindent 
We thank Lucio Isa, Ryohei Seto, and Mingcheng Yang for their insightful discussions. This work was supported by the General Research Fund (No. 16307422) and the Collaborative Research Fund (No. C6004-22Y) from the Hong Kong Research Grants Council (RGC). We also acknowledge the support of the HKUST Marine Robotics and Blue Economy Technology Grant (No. MRBET-002). Y.Z. acknowledges funding from the RGC Postdoctoral Fellowship (No. PDFS2324-6S02). Y.Z. and C.X. further thank the Yukawa Institute for Theoretical Physics (YITP) at Kyoto University for its warm hospitality and the productive discussions during the Frontiers in Non-equilibrium Physics 2024 workshop (YITP-T-24-01).

\vspace{11pt}
\noindent{\bf  Author Contributions}

\noindent
C.~X., Y.~Z., W.~H. and Q.~X. designed the project. C.~X., S.~W., H.~W., X.~Liu, and Z.~L. conducted the experimental measurements. C.~X., S.~W., H.~W., Y.~Z., and Q. X. analyzed the experimental data. S.~W., C.~X., Z.~L., and Q.~X. conducted the simulations. C.~X, S.~W., Y.~Z., W.~H., and Q.~X. wrote the manuscript. 

\vspace{11pt}
\noindent {\bf  Competing interests} 

\noindent
The authors declare that they have no conflict of interest.

\vspace{11pt}
\noindent {\bf  Data and materials availability}

\noindent
The data that support the findings of this study are available from the authors on request.

\begin{figure*}
        \addtocounter{figure}{-5} 
\centering
    \begin{figurehereExtend}
    \centering
\includegraphics[width=15cm]{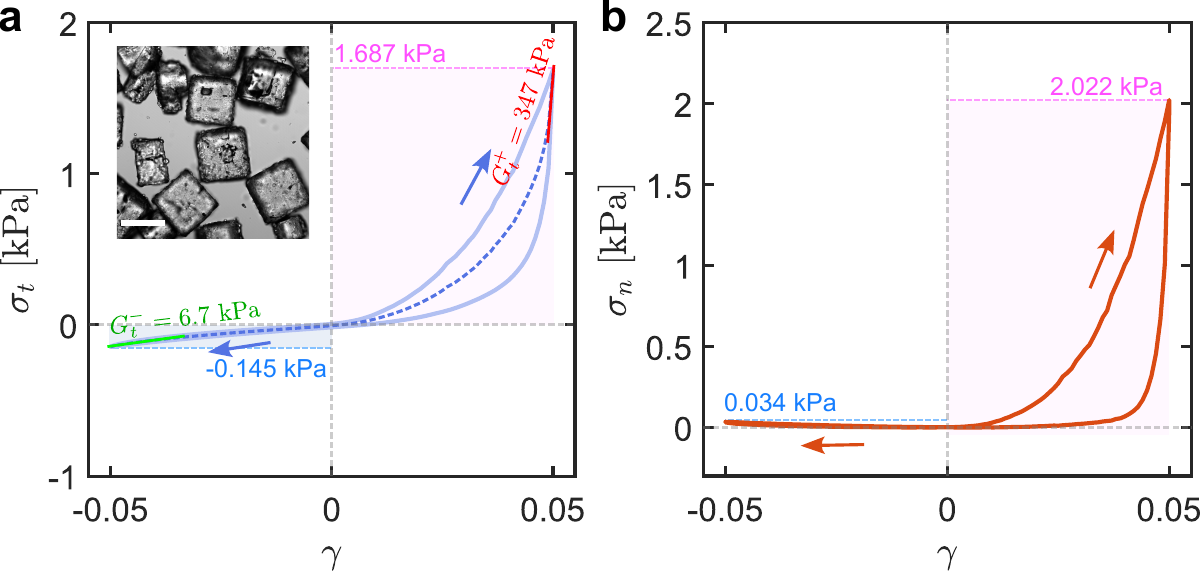}
\caption{ \textbf{Enhanced non-reciprocal responses of NaCl–PDMS composites.} (a) Plot of asymmetric mechanical responses in shear stress ($\sigma_t$) against shear strain ($\gamma$) for NaCl–PDMS composites ($\phi = 59\%$, $G_m = 0.11$~kPa, and $\sigma_t^* = 80$~Pa). The dashed line indicates the averaged stress-strain curve. The differential moduli $G_t^+$ and $G_t^-$ are obtained by measuring the slopes of the averaged stress-strain curve at $\gamma$=0.05 and $\gamma$=-0.05. At $\gamma = \pm 0.05$, the degree of non-reciprocity in shear stress and differential modulus reaches $\vert \sigma_t^+/\sigma_t^-\vert \approx 10$ and $G_t^+/G_t^- \approx 50 $, respectively. Inset: Image of NaCl hexahedral particles in uncured PDMS base solution. Scale bar: 300 $\mu$m. (b) Plot of asymmetric normal stress ($\sigma_n$) against shear strain ($\gamma$).}
    \label{fig:Enhanced}
\end{figurehereExtend}
\end{figure*}

\bibliography{reference}

\end{document}